\documentclass[journal]{IEEEtran}

\usepackage[latin1]{inputenc}
\usepackage{times}
\usepackage{amsmath}
\usepackage{amssymb}
\usepackage{mathrsfs}
\usepackage{subcaption}
\usepackage{graphicx}
\usepackage{enumerate}
\usepackage[compress]{cite}
\usepackage{float}

\begin{document}

\title{Statistical Analysis and Optimization of a 5th-Percentile User Rate Constrained Design for FFR/SFR-aided OFDMA-based Cellular Networks}

\author{Jan~Garc\'ia-Morales,~\IEEEmembership{Student Member,~IEEE,}
        Guillem~Femenias,~\IEEEmembership{Senior Member,~IEEE,}
        and~Felip~Riera-Palou,~\IEEEmembership{Senior Member,~IEEE}
\thanks{J. Garc\'ia-Morales, G Femenias and F Riera-Palou are with the Mobile Communications Group, University of the Balearic Islands, Palma 07122, Illes Balears, Spain (e-mail: \{jan.garcia,guillem.femenias,felip.riera\}@uib.es).}
%\thanks{Manuscript received September 15, 2016.}
}

%\markboth{Submitted to IEEE Transactions on Vehicular Technology}{Garc\'ia-Morales %\textit{et al.}: Statistical Analysis and Optimization of a 5th-Percentile User Rate Constrained Design for FFR/SFR-aided OFDMA-based Cellular Networks}

\maketitle

\begin{abstract}
Interference mitigation strategies are deemed to play a key role in the context of the next generation (B4G/5G) of multi-cellular networks based on orthogonal frequency division multiple access (OFDMA). Fractional and soft frequency reuse (FFR, SFR) constitute two powerful mechanisms for inter-cell interference coordination (ICIC) that have been already adopted by emerging cellular deployments as an efficient way to improve the throughput performance perceived by cell-edge users. This paper presents a novel optimal 5th-percentile user rate constrained design (R5pD) for FFR/SFR-based networks that, by appropriately dimensioning the center and edge regions of the cell, rightly splitting the available bandwidth among these two areas while assigning the corresponding transmit power, allows a tradeoff between cell throughput performance and fairness to be established. To this end, both the cumulative distribution function of the user throughput and the average spectral efficiency of the system are derived assuming the use of the ubiquitous proportional fair scheduling policy. The analytical framework is then used to obtain numerical results showing that the novel proposed design clearly outperforms previous schemes in terms of throughput fairness control due to a more rational compromise between average cell throughput and cell-edge ICIC.
\end{abstract}


% \fi
\IEEEpeerreviewmaketitle

\section{Introduction}
\IEEEPARstart{M}{odern} cellular communications standards such as Long-Term Evolution (LTE) and LTE-Advanced (LTE-A) have adopted the orthogonal frequency division multiple access (OFDMA) as the downlink radio interface \cite{Dahlman13}, \cite{bhat2012lte}. Furthermore, it is envisaged that OFDMA will keep playing a major role in 5G cellular networks \cite{andrews2014will}. The physical layer of an OFDMA network relies on the division of a wideband frequency-selective channel into multiple orthogonal narrowband frequency-flat channels that, when operated in a time-slot manner, define a time/frequency grid whose elements (resource blocks (RBs)) serve as the basic resource allocation unit. Owing to the RB orthogonality, no intracell interference arises, however, since most cellular deployments nowadays rely on aggressive universal frequency reuse schemes, users situated in the cell-edge areas are prone to experience large levels of inter-cell interference (ICI). In order to address the performance degradation ICI brings along while still pursuing high spectral efficiencies, a plethora of ICI coordination (ICIC) techniques have been investigated \cite{Hamza13}, among which fractional and soft frequency reuse (FFR, SFR) and variants thereof, have achieved widespread use thanks to the excellent compromise they provide between overall spectral efficiency and cell-edge performance.

In FFR-based cellular systems a low frequency reuse factor is used for the cell-center users, less affected by ICI, and a larger frequency reuse factor is selected for the cell-edge users, much more exposed to ICI. SFR is a variation of FFR whereby the central region of each cell is also allowed to employ the frequency resources allocated to the edges of the neighboring cells, thus the whole system bandwidth is reused in every cell \cite{huawei05}. Furthermore, different transmit powers can be allocated to the central and edge RBs in SFR-based cellular systems, with the aim of increasing the signal-to-interference-plus-noise ratio (SINR) experienced by those users located far away from the BS. Summarizing, FFR/SFR-based frequency reuse schemes provide compound reuse factors usually lying between 1 and 3, thus sacrificing spectral efficiency in favor of some degree of fairness between cell-center and cell-edge users. One of the major issues when dealing with FFR/SFR-based ICI mitigation strategies is the design of long-term resource allocation algorithms aiming at the optimization of throughput-related utility functions with constraints on the degree of fairness between users located throughout the cell \cite{andrews2014will,garciaperformanceMONET}. In order to tackle these optimization problems, it is very convenient to obtain mathematically tractable analytical models for both the system and user throughput. Unfortunately, the analytical throughput models used to design the long-term FFR/SFR-based resource allocation algorithms are fully related to the short-term scheduling rules used to select the set of users on each time/frequency/energy resource, thus making the derivation of these analytical models a challenging task.

It is well known that, regardless of the ICIC technique employed, channel-aware schedulers are able to exploit the inherent system multiuser diversity by allocating each RB, on a slot-by-slot basis, to users enjoying \emph{favourable} channel conditions (i.e., users experiencing high SINR values), and thus leading to a significant enhancement of network spectral efficiency. Amongst many scheduling rules, the proportional fair (PF) scheduler \cite{Kelly98} has been successfully deployed in the latest generations of wireless networks, owing to the excellent trade-off it offers between fairness and spectral efficiency. Hence, the obtention of exact mathematically tractable analytical models for the system and user throughput of FFR/SFR-aided OFDMA-based cellular networks under PF scheduling is a challenging aim of significant interest when designing optimal long-term resource allocation strategies.

\subsection{Background work}
Several research studies have focused on the theory of stochastic geometry, where the BSs are distributed using Poisson Point Processes (PPP) \cite{Novlan11,ElSawy13,Kumar15a} or hard-core PPPs \cite{Heath13}. However, the use of stochastic geometry to characterize cellular layouts hinders to accurately model the consequences of using ICIC techniques such as FFR/SFR when considering regularly-planned cellular arrangements.

In contrast to the above background work, which relies on the use of stochastic geometry to model the cellular environment, the work in \cite{Jin13} does indeed account for the regular deployment using an FFR-aided OFDMA-based network. Unfortunately, the analytical framework proposed by Fan Jin \emph{et al.} in \cite{Jin13} is limited to the use of resource allocation schemes based on round robin (RR) scheduling. In \cite{boddu15}, the authors investigate the cell capacity and the cell-edge performance in terms of user satisfaction. They present an extensive comparison between FFR and SFR, but using again the RR scheduling rule. Similar approaches, lacking the consideration of channel-aware scheduling policies, are also proposed by Assaad in \cite{Assaad08}, Najjar \emph{et al.} in \cite{Najjar09}, Kumar \emph{et al.} in \cite{Kumar15b}, Hambebo \emph{et al.} \cite{Hambebo14} and Hung-Bin Chang \emph{et al.} in \cite{Chang16} to optimize FFR-based network parameters. A generalized unifying analytical framework suitable for both FFR and SFR while also considering the uplink has been recently presented in \cite{Mahmud14} but again, as the previous aforementioned works, neglecting the role of the scheduler. Closer to the problem addressed in this manuscript, Sagkriotis \emph{et al.} in \cite{Sagkriotis16} target the optimization of an FFR system in terms of maximum capacity and power efficiency when assuming the use of a PF scheduler. However, the proposed design does not incorporate any constraints on the cell-edge performance and thus, it cannot provide QoS guarantees to the \emph{worst} users in the network.

Some of the limitations of this background work have been partially overcome in our contributions \cite{garciaperformanceMONET} and \cite{garciaanalysisAccess}, where an analytical framework was developed that allows the evaluation of the cell throughput when using opportunistic maximum SINR (MSINR) and PF schedulers. However, our previous works considered only the case of \emph{strict} FFR and, moreover, the proposed designs were based on average cell-center and cell-edge metrics, thus neglecting a complete statistical characterization of the throughput \cite{bhat2012lte}. Similarly, \cite{Zhu14} presents an analytical performance model for FFR when performing chunk-based (rather than subcarrier-based) resource allocation but again focusing on the maximization of the average spectral efficiency and thus neglecting any minimum service guarantees to the worst users in the network. More recently Wang et al. in \cite{Wang16} have proposed an extension of the FFR scheme by allowing more than one reuse outer layers in an attempt to improve the performance on cell-edge region but their design does not consider the incorporation of explicit QoS requirements for the users in this region and furthermore, the effects of fast fading and the use of channel-aware schedulers have both been neglected. Remarkably, most 5G defining proposals (e.g., \cite{ElAyoubi16}) specify key performance indicators (KPIs) whose targets must be accomplished by at least a certain percentage of users, regardless of their position. To this end, a full statistical characterization of the throughput is required. In \cite{garcia2016characterizing}, we presented some preliminary results of our effort to close this gap. In that work, we proposed an analytical framework allowing the obtention of statistical models of the average user throughput, but we did not consider the optimal designs required to maximize the cell throughput while guarantying a certain degree of quality-of-service (QoS) throughout the cell.

We note at this point, that the focus of this work is on single-tier networks. Extending the analytical framework proposed here to heterogeneous setups by combining results presented in \cite{garciaanalysisAccess} with multi-tier specific ICIC techniques such as those based on almost blank subframes \cite{Deb14,Zhou17,Zheng17}, constitutes an interesting avenue for further research.

\subsection{Contributions of the paper}
In this paper, different optimal designs for a downlink OFDMA-based multi-cellular network are studied and compared. To this end, an analytical framework is presented allowing the performance evaluation of both SFR and FFR schemes using a PF scheduling policy. The main contributions of this paper can be summarized as follows:
\begin{itemize}
    \item A novel analytical optimization approach, which we term 5th-percentile user rate constrained design (R5pD), is introduced whereby a cell throughput-based utility function maximization is conducted subject to a constrained 5th-percentile user rate\footnote{It should be noted that the focus of this paper is on the 5th-percentile since that is the percentage typically specified in most curent standards and future proposals. Nevertheless, the framework developed here is generally applicable to any $x$th-percentile constrained design.}. Guarantying a specified value for the 5th-percentile user rate is one of the most important technical challenges in current and emerging cellular systems trying to provide uniform QoS levels throughout the cell \cite{bhat2012lte}.

   \item In order to deal with the R5pD-based design, a global characterization of throughput is introduced for both SFR- and FFR-aided OFDMA-based cellular networks using PF scheduling. This analytical framework is used to derive mathematically tractable expressions for important metrics such as the cumulative distribution function (CDF) of the average user throughput and the average system spectral efficiency.

   \item Furthermore, results for the R5pD-based strategy are compared to those obtained using well-known optimal designs, namely, the fixed spectrum/power factor design (FxD) and the QoS-constrained design (QoScD), whose optimization criteria strike a balance between the cell-edge and cell-center throughput leading to different network performance behaviors. Moreover, the proposed framework allows an exhaustive comparison between FFR and SFR to be carried out that reveals under what conditions one frequency reuse technique is preferable over the other.
\end{itemize}
Although results are obtained for an FFR/SFR-aided deployment, this analytical framework opens the door to the theoretical spectral efficiency evaluation of OFDMA-based cellular networks using more sophisticated ICIC techniques such as adaptive frequency reuse or network MIMO, as well as to the assessment of cellular multi-tier networks where the macro-cellular network is underlaid by different tiers of pico- and femto-cellular BSs \cite{Andrews13,garciaanalysisAccess}.

The rest of the paper is organized as follows. In Section \ref{sec:System_model} the system model under consideration in the context of FFR/SFR-aided OFDMA-based networks is introduced. The novel optimal R5pD-based design is presented in Section \ref{sec:Optimization}. The statistical characterization of both the users throughput and the average cell throughput are derived in Section \ref{sec:Throughput}. Extensive analytical and simulation results, for each reuse scheme and using the PF scheduler, are provided in Section \ref{sec:Results}. Finally, the main outcomes of this paper are recapped in Section \ref{sec:Conclusion}.

\section{FFR/SFR system model}
\label{sec:System_model}

\subsection{Network model}

\begin{figure}
\centering
\includegraphics[width=0.45\textwidth]{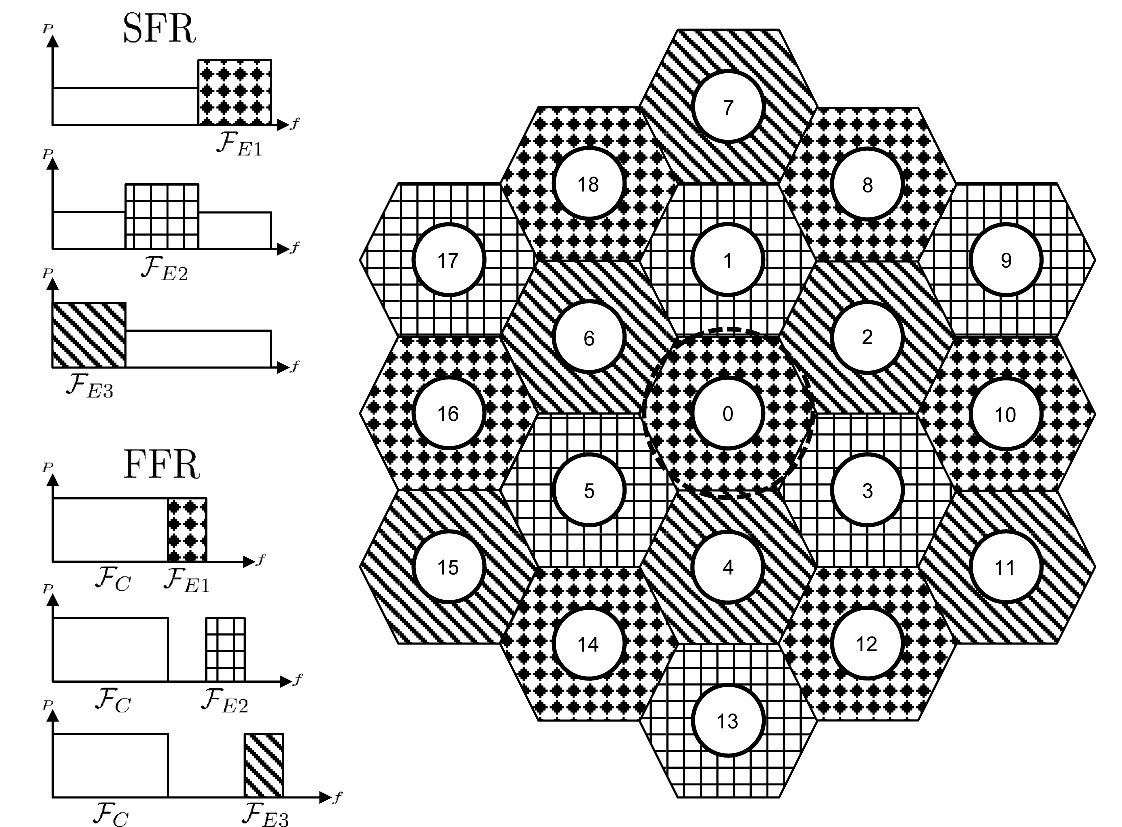}
\caption{FFR/SFR-aided OFDMA network topology.}
\label{fig:Network_model}
\end{figure}

This work focuses on the downlink of an FFR/SFR-aided OFDMA-based multi-cellular system that, as shown in Fig. \ref{fig:Network_model}, has been modeled as a regular tessellation of hexagonal cells whose serving BSs are located at their centers\footnote{This paper assumes the use of omnidirectional antennas at the BSs leaving for further work the issues related to antenna sectorization.}. Towards the implementation of FFR/SFR, a user categorization is first conducted. In particular, users whose average SINR is above a prescribed threshold $\Gamma_{th}$ are deemed as central ones, whereas the rest, those whose average SINR is below $\Gamma_{th}$, are considered to be edge users. Subsequently,
non-overlapping frequency bands are allocated to cell-center and cell-edge users. The frequency bands are usually designed in such a way that a low frequency reuse factor can be used to serve the central users and a higher reuse factor can be used for the edge users. Throughout this paper, and without loss of generality, a frequency reuse factor of 3 is assumed for cell-edge users whereas central users are assumed to employ a reuse factor of 1 (universal). Furthermore, and for the sake of analytical tractability, it is also assumed that a circumference of radius $R_{th}$ (threshold distance) delimitates the cell-center and cell-edge regions. Finally, the central cell constituting the cell/BS of interest is approximated by a circle of the same area as the corresponding hexagon (see Fig. 1). This implies that if the side of the regular hexagon is denoted by $R_h$, the radius of the circular cell is $R_m=R_h\sqrt{3\sqrt{3}/2\pi}$.

The full system bandwidth is exploited by means of a set $\mathcal{F}_T$ of $N_{\text{RB}}$ RBs (or orthogonal subbands), each made of $N_{sc}$ adjacent subcarriers and whose bandwidth, $B_{\text{RB}}$, is small enough so that all subcarriers in a subband can be safely assumed to experience frequency flat fading. When using FFR, the set $\mathcal{F}_T$ is divided into a subset $\mathcal{F}_C$ of cell-center RBs and a subset $\mathcal{F}_T \backslash \mathcal{F}_C$ of cell-edge RBs. We define the number of RBs allocated to the cell-center area to be $N_C=\rho N_{\text{RB}}$, where $0{\leq\rho\leq}1$ is the spectrum allocation factor. The set $\mathcal{F}_T \backslash \mathcal{F}_C$ is then segmented into three equal parts, namely $\mathcal{F}_{E1}$, $\mathcal{F}_{E2}$ and $\mathcal{F}_{E3}$, of size $N_E=(1-\rho)N_{\text{RB}}/3$, that are assigned to cell-edge users while taking care that adjacent cells will use different subsets of RBs. Note that $N_C=N_{\text{RB}}-3 N_E$ must be a non-negative integer value less or equal than $N_{\text{RB}}$ and thus, $N_E\in\{0,1,\ldots,\lfloor N_{\text{RB}}/3\rfloor\}$ and $\rho$ can only take values in the set
\begin{equation}
   \mathcal{S}_\rho=\left\{\frac{N_{\text{RB}}-3\lfloor N_{\text{RB}}/3\rfloor}{N_{\text{RB}}}, \frac{N_{\text{RB}}-3(\lfloor N_{\text{RB}}/3\rfloor-1)}{N_{\text{RB}}},\ldots,1\right\},
   \label{eq:Srho}
\end{equation}
where $\lfloor x\rfloor$ denotes the floor operator.
When relying on  SFR, the set $\mathcal{F}_T$ is divided into three subsets of $N_{\text{RB}}/3$ RBs that, with a slight abuse of notation, will also be denoted by $\mathcal{F}_{E1}$, $\mathcal{F}_{E2}$ and $\mathcal{F}_{E3}$. On any given cell, one of the RB subsets is allocated to the cell-edge area, thus $N_E=N_{\text{RB}}/3$, while the remaining RBs are assigned to the cell-center area, that is, $N_C = 2N_{\text{RB}}/3$.

The positions of the users at a given instant are assumed to conform to a stationary PPP of normalized intensity $\lambda$ (measured in users per area unit). As a consequence, the probability distribution of the number $M_\mathcal{S}$ of users falling within any spatial region $\mathcal{S}$ of area $A_\mathcal{S}$ adheres to a Poisson distribution, thus implying
\begin{equation}\label{e0}
Pr\{M_\mathcal{S}=k\} = \frac{(\lambda A_\mathcal{S})^k e^{-\lambda A_\mathcal{S}}}{k!}.
\end{equation}

\subsection{Channel model}

The downlink channel includes both a distance-dependent path loss and a small-scale fading\footnote{As it is typically done in the background literature (see \cite{Novlan11,Jin13,Xu12,Femenias15}), and for analytical simplicity, only pathloss and small scale fading are taken into account in this paper while leaving the incorporation of large scale fading (shadowing) to future work. Note that in this case, as the cellular networks under consideration are always limited by inter-cell interference, average SINR- and distance-based strategies used to classify users as cell-center or cell-edge are virtually equivalent.}. The instantaneous SINR experienced by user $u$ in the cell of interest, also denoted as cell 0, on any of the $N_{sc}$ subcarriers belonging to the $n$th RB during the period $t$ can be expressed as
\begin{equation}\label{e2}
   \gamma_{u,n}(t) = \frac{ P_n L(d_{0,u})|H_{0,u,n}(t)|^2}{N_0 \Delta f + I_{u,n}(t)},
\end{equation}
where, under the assumption of uniform power allocation, $P_n$ is the power allocated per subcarrier, $d_{b,u}$ represents the distance between BS $b$ and user $u$, $L(d_{b,u})$ is the path loss component characterizing the link between the $b$th BS and the $u$th user, $H_{b,u,n}(t)\sim \mathcal{C N} (0,1)$ is the frequency response\footnote{Note that, since the proposed scheduling and resource allocation framework are rooted on capacity formulas optimized on an RB-by-RB basis, the possible frequency correlation among RBs is irrelevant, thus making unnecessary the specification of any frequency correlation profile.} resulting from the small-scale fading channel linking the $b$th BS to user $u$ on the $n$th RB during scheduling period $t$, $N_0$ is the noise power spectral density, $\Delta f=B_{\text{RB}}/N_{sc}$ is the subcarrier bandwidth, and $I_{u,n}(t)$ corresponds to the interference term.

In the particular case of FFR, the ICI can be obtained as
\begin{equation}\label{e3}
   I_{u,n}(t)=\sum_{b\in \Phi_n} P_n L\left(d_{b,u}\right)|H_{b,u,n}(t)|^2,
\end{equation}
where $P_n=P_T/(N_{sc}(N_C+N_E))$, with $P_T$ denoting the transmit power available at the BS, and $\Phi_n$ represents the RB-dependent set of interfering BSs.
In the SFR configuration, the transmit power allocated per cell-edge subcarrier $P_n^E=P_T/(N_{sc}(\beta N_C+N_E))$ is larger than the corresponding one allocated per cell-center subcarrier $P_n^C=P_T/(N_{sc}(N_C+N_E/\beta))$, where $P_n^C = \beta P_n^E$, with $0{\leq\beta\leq}1$ denoting the power control factor. Hence, in \eqref{e2} it holds that $P_{n}= P_n^E$ for a cell-edge user and $P_{n}= P_n^C$ for a cell-center user. Hence, the SFR interference term is given by
\begin{equation}\label{e4}
\begin{split}
   I_{u,n}(t)=&\sum_{b\in \Phi_n^C} P_n^C L\left(d_{b,u}\right)|H_{b,u,n}(t)|^2 \\
             +&\sum_{b\in \Phi_n^E} P_n^E L\left(d_{b,u}\right)|H_{b,u,n}(t)|^2\textrm{ },
\end{split}
\end{equation}
where $\Phi_n^C$ and $\Phi_n^E$ denote the sets of interfering BSs affecting cell-center and cell-edge users, respectively.

\section{Optimal R5pD-based design}
\label{sec:Optimization}
In this section we propose a 5th-percentile user rate constrained design for an FFR/SFR-aided OFDMA-based multi-cellular network. The proposed optimization strategy aims at determining the size of the reuse scheme-related spatial, frequency and power partitions maximizing the average cell throughput while satisfying the corresponding operator-defined system constraints.
The parameters used to pose this novel optimal design are the distance threshold ratio $\omega \triangleq R_{th}/R_m$, the spectrum allocation factor $\rho \triangleq N_C/N_{\text{RB}}$ and the power control factor $\beta \triangleq P_n^C/P_n^E$. The appropriate selection of these parameters significantly affects the average cell throughput (measured in bps) that can be expressed as
\begin{equation}\label{d0}
    \overline{\eta}(\omega,\zeta) = \overline{\eta}^C(\omega,\zeta) + \overline{\eta}^E(\omega,\zeta),
\end{equation}
where $\overline{\eta}^A(\omega,\zeta)$ represents the average throughput in the cell-region $A$, being $A$ a token used to represent either the cell-center region $C$ or the cell-edge region $E$ and $\zeta$ represents a token that can be either $\rho$ when using FFR or $\beta$ using SFR.

Under R5pD, the constraint of the 5th-percentile user rate $R5\%$ is used. This is an important fairness metric to consider, since modern cellular networks are increasingly required to provide high data-rate and also guaranteed QoS throughout the cell. In fact, 5G initiatives explicitly state requirements for the 5th-percentile rate \cite{andrews2014will}. The $R5\%$ (measured in bps) reflects the throughput achieved by the worst users in the system, typically located at the cell-edge. In general, the $x$th-percentile of users rate $Rx\%$  is defined as the $x$\%-tile of the users throughput CDF, that is, $F_{\overline{\eta}_u}(Rx\%)=x/100$. The optimization problem using R5pD can then be formulated as
\begin{equation}\label{d1}
\begin{split}
(\omega^*,\zeta^*) = & \arg\max_{\substack{\frac{R_{0m}}{R_m}{\leq\omega\leq}1 \\ 0 \leq \zeta \leq 1}} \overline{\eta}^C(\omega,\zeta) + \overline{\eta}^E(\omega,\zeta),\\
                     & \textup{ subject to } F^{-1}_{\overline{\eta}_u}(0.05) \geq v_o
\end{split}
\end{equation}
where $R_{0m}$ denotes the minimum distance of a user from its serving BS and $v_o$ represents the throughput requirement.

Hence, using R5pD allows the control of a more equitable distribution of both spectrum and transmit power that can result in a higher degree of fairness between cell-center and cell-edge users, at the cost of reducing the average cell throughput with respect to the well-known FxD-based design \cite{Xu12,Femenias15,garciaperformanceMONET}.
It is worth pointing out that the optimization problem \eqref{d1} can be efficiently solved using standard software optimization packages (e.g., Matlab).
In following sections the analytical framework allowing the theoretical calculation of the performance metrics used to pose the optimization problem in \eqref{d1} is developed.

\section{Characterization of the throughput}
\label{sec:Throughput}

\subsection{Statistical characterization of the users throughput}
\label{sec:Throughput_CDF}
By definition, the CDF characterizing the average throughput allocated to user $u$ in region $A$ can be expressed as $F_{\overline{\eta}^A_u}(v)=Pr\{\overline{\eta}_u^A \leq v\}$. Due to the Poissonian distribution of users in a cell, the number of users in the cell of interest, denoted as $M_0$, is a non-negative integer random variable. For a particular event $M_0=k$, with $k\in\{0, 1, \ldots,\infty\}$, the random variable $k_A$, representing the number of users in cell region $A$, can only take values in the set $\{0,1,\ldots,k\}$. Furthermore, for a particular value of $k_A$ there are a total of ${k\choose k_A}$ combinations of users in which $k_A$ of them are located in region $A$ and the remaining $(k-k_A)$ are located in the tagged cell but outside region $A$. As all previous events are mutually exclusive and its probabilities sum to unity, the total probability theorem can be applied to obtain
\begin{equation}\label{eCDFr1}
\begin{split}
F_{\overline{\eta}^A_u}(v)&=\sum_{k=1}^{\infty} Pr\{M_0=k\} \sum_{k_A=1}^k {k\choose k_A} P_A^{k_A}\\
                          &\times \left(1 - P_A\right)^{k-k_A} \left[\frac{k_A}{k} Pr\{\overline{\eta}_u^A(k_A) \leq v | k_A\}\right],
\end{split}
\end{equation}
where $P_A$ is the probability that a user is located in cell-region $A$ and $\overline{\eta}_u^A(k_A)$ represents the average throughput experienced by a user $u$ located in cell-region $A$ conditioned on the event that there are $k_A$ users in that region. As users are assumed to be uniformly distributed in the cell, then
\begin{equation}\label{T2}
  P_A=\frac{(R_U^A)^2-(R_L^A)^2}{R_m^2 - R_{0m}^2},
\end{equation}
where $R_L^A$ and $R_U^A$ are the lower and upper radii of the circumferences delimiting cell-region $A$.

Now, defining $M_A$ as a non-negative integer random variable representing the number of users in the cell-region $A$, the average throughput experienced by a user $u$ located in cell-region $A$ when $M_A = k$, can be expressed as
\begin{equation}\label{T3}
  \begin{split}
    \overline{\eta}^A_u(k) &= \mathbb{E}_{d_{0,u}} \left\lbrace N_A \textup{ } \overline{\eta}^A_{u,n}(k,d)\right\rbrace \\
                           &= N_A \int_{R_L^A}^{R_U^A} \overline{\eta}^A_{u,n}(k,d)  f_{d_{0,u}}(d)\textrm{ }\mathrm{d}d,
  \end{split}
\end{equation}
where $\overline{\eta}_{u,n}^A(k,d)$ is the scheduler-dependent average throughput experienced by a user $u$ on the $n$th RB in cell-region $A$, $N_A$ is the number of RBs allocated to cell-region $A$ and $f_{d_{0,u}}(d)$ is the probability density function (PDF) of the random variable $d_{0,u}$.

Relying on the knowledge of the instantaneous SINRs experienced by all users $q\in\mathcal{M}_A$, the PF scheduler allocates RB $n\in\mathcal{F}_A$ to the user $u\in\mathcal{M}_A$ satisfying
\begin{equation}
   u = \arg\max_{q\in\mathcal{M}_A}\{w_q(t) \gamma^A_{q,n}(t)\},
   \label{PF1}
\end{equation}
where $\mathcal{M}_A$ is the set used to index all users in cell-region $A$, and $w_q(t)=1/\mu_q(t)$ is the weighting (prioritization) coefficient for user $q$, which can be computed using a moving average of the short-term CSI over a window of $W$ scheduling periods as
\begin{equation}
   \mu_q(t)=\left(1-\frac{1}{W}\right)\mu_q(t-1)+\sum_{n\in\mathcal{F}_A} \iota_{q,n}(t)\frac{\gamma^A_{q,n}(t)}{W},
   \label{PF2}
\end{equation}
with $\iota_{q,n}(t)$ being the indicator function of the event that user $q$ during scheduling period $t$ is scheduled on RB $n$.

It is shown in \cite{Liu10} that, for large values of $W$ and after the PF scheduler becomes stable, $\mu_q(t)$ exhibits very little variation with $t$ and thus, it can be harmlessly approximated by its statistical expectation\footnote{In particular, and as it has been duly checked via simulation results, the approximation of constant PF weights is valid for window lengths as low as $W=50$ scheduling periods.}, that is, $\mu_q(t)\simeq\mathbb{E}\{\mu_q(t)\}\triangleq \overline{\mu}_q$. Hence, user $u \in \mathcal{M}_A$ will be scheduled on subcarrier $n\in\mathcal{F}_A$ whenever
\begin{equation}
   \varphi^A_{u,n}(t) > \varphi^A_{\max,u,n}(t) \triangleq \max_{\substack{q\in\mathcal{M}_A \\ q \neq u}} \left\{\varphi^A_{q,n}(t)\right\},
\end{equation}
where $\varphi^A_{q,n}(t)\triangleq \gamma^A_{q,n}(t)/\overline{\mu}_{q}$.
According to the previous definition of the PF scheduler, the instantaneous throughput experienced by a user $u$ on the $n$th RB in cell-region $A$, conditioned on the event that there are $M_A=k$ users and on the set of distances $\boldsymbol{d}=\{d_{0,u}\}_{\forall\,u\in\mathcal{M}_A}$, can be evaluated as\footnote{Note that since the channel is assumed to be stationary, from this point onwards the time dependence (i.e., (t)) of
all the variables will be dropped unless otherwise stated.}
\begin{equation}\label{ePF_Thr1}
\begin{split}
&\eta_{u,n}^A(\gamma_{u,n}|k,\boldsymbol{d}) \\
&\qquad=\begin{cases}
           \qquad\qquad 0, & \varphi^A_{\max,u,n}>\frac{\gamma^A_{u,n}}{\overline{\mu}_u}\\
           B_{\text{RB}} \log_2(1 + \gamma^A_{u,n}), & \varphi^A_{\max,u,n}\leq\frac{\gamma^A_{u,n}}{\overline{\mu}_u}
        \end{cases}
\end{split}
\end{equation}
and the average throughput experienced by a user $u$ on the $n$th RB in cell-region $A$ can be obtained as
\begin{equation}\label{ePF_Thr2}
\begin{split}
\overline{\eta}_{u,n}^A(k,\boldsymbol{d}) &= B_{\text{RB}} \int_0^{\infty} \log_2(1+x) f_{\gamma^A_{u,n}|d_{0,u}} \left(x|d\right)\\
&\quad\qquad\times F_{\varphi^A_{\max,u,n} | \boldsymbol{d}}\left(\frac{x}{\overline{\mu}_u} \,\Bigr|\,\boldsymbol{d}\right)\mathrm{d}x\\
&=B_{\text{RB}} \int_0^{\infty} \log_2(1+x) f_{\gamma^A_{u,n}|d_{0,u}} \left(x|d\right)\\
&\quad\qquad\times \prod_{\substack{q\in\mathcal{M}_A \\ q \neq u}} F_{\varphi^A_{q,n} | d_{0,q}} \left(\frac{x}{\overline{\mu}_u} \,\Bigr|\,d_{0,q} \right) \mathrm{d} x,
  \end{split}
\end{equation}
where $F_{\varphi^A_{\max,q,n} | \boldsymbol{d}}\left(x |\boldsymbol{d}\right)$ is the conditional CDF of $\varphi^A_{\max,q,n}$ conditioned on the set of distances $\boldsymbol{d}$, and $f^A_{\gamma_{q,n}|d_{0,q}}(x | d_{0,q})$ and $F^A_{\varphi_{q,n}|d_{0,q}}(x | d_{0,q})$ are used to denote, respectively, the PDF of $\gamma^A_{q,n}$ and the CDF of $\varphi^A_{q,n}$ conditioned on $d_{0,q}$.

Under the large $W$ assumption, $\overline{\mu}_q$ tends to be proportional to $\mathbb{E}\{\gamma^A_{q,n}(t)\}$ and thus, the random variable $\varphi^A_{q,n}(t)= \gamma^A_{q,n}(t)/\overline{\mu}_{q}$, as shown by Liu and Leung in \cite{Liu10}, tends to be unrelated to the pathloss component. Consequently, it can be safely assumed that the conditional random variables $\{\varphi^A_{q,n}|d_{0,q}\}_{\forall\,q\in\mathcal{M}_A}$ are independent and identically distributed (i.i.d.). That is, given the positions of users in region $A$, it is assumed that on each subcarrier $n$ in region $A$ the users are statistically equivalent in terms of the scheduling metrics. Then expression \eqref{ePF_Thr2} simplifies to
\begin{equation}\label{eq:PF_CDF}
    \begin{split}
        \overline{\eta}_{u,n}^A(k,d) &= B_{\text{RB}} \int_0^{\infty} \log_2(1+x) f_{\gamma^A_{u,n}|d_{0,u}} \left(x|d\right)\\
        &\quad\qquad\times \left(F_{\gamma^A_{u,n}|d_{0,u}}(x|d)\right)^{k-1} \mathrm{d}x,
    \end{split}
\end{equation}
where the conditional CDF $F_{\gamma^A_{u,n}|d_{0,u}}(x|d)$ of the instantaneous SINR experienced by user $u$ on any subcarrier of the $n$th RB can be obtained as shown in the Appendix.

Let us now define the distance $d(k,v) \triangleq \{d:\overline{\eta}_u^A(k,d)=v\}$. Since $\overline{\eta}_u^A(k,d)$ is a decreasing function of $d$ on the interval $\left[R_L^A,R_U^A\right]$, then
\begin{equation}\label{eCDFr2}
\begin{split}
Pr&\{\overline{\eta}_u^A(k) \leq v|k\}=Pr\{d_{0,u} \geq d(k,v)\}\\
            &=\left\lbrace \begin{split}
                           & \textup{ }\textup{ }\textup{ }\textup{ }\textup{ }\textup{ } 1 \textup{ }\textup{ }\textup{ }\textup{ }\textup{ }\textup{ }\textup{ }\textup{ }\textup{ }\textup{ }\textup{ }\textup{ },\textup{ } d(k,v) \leq R_L^A\\
                           & \frac{(R_U^A)^2 - {d(k,v)}^2}{(R_U^A)^2-(R_L^A)^2},\textup{ } R_L^A \leq d(k,v) \leq R_U^A\\
                           & \textup{ }\textup{ }\textup{ }\textup{ }\textup{ }\textup{ } 0 \textup{ }\textup{ }\textup{ }\textup{ }\textup{ }\textup{ }\textup{ }\textup{ }\textup{ }\textup{ }\textup{ }\textup{ },\textup{ } d(k,v) \geq R_U^A
            \end{split} \right..
\end{split}
\end{equation}
Thus, substituting \eqref{eCDFr2} in \eqref{eCDFr1}, the users throughput CDF of FFR/SFR-aided OFDMA-based multi-cellular networks can be finally expressed as
\begin{equation}\label{eCDFr4}
        F_{\overline{\eta}_u}(v) \triangleq Pr\{\overline{\eta}_u \leq v\} = F_{\overline{\eta}^C_u}(v) + F_{\overline{\eta}^E_u}(v).
\end{equation}

\subsection{Cell throughput analysis}
\label{sec:Cell_Throughput}
The downlink average throughput experienced in the cell-region $A$ of the FFR/SFR-aided OFDMA-based multi-cellular network can be derived as
\begin{equation}\label{T1}
  \begin{split}
    \overline{\eta}^A(\omega,\zeta) =& \sum_{k=1}^{\infty} Pr\{M_0=k\} \sum_{k_A=1}^k {k\choose k_A} P_A^{k_A}  \\
    &\times \left(1 - P_A\right)^{k-k_A} \left[k_A \textup{ } \overline{\eta}_u^A(k_A)\right].
  \end{split}
\end{equation}

Substituting the expressions \eqref{eq:PF_CDF} and \eqref{T3} in \eqref{T1}, the average throughput in the cell-region A can be simplified to
\begin{equation}
\begin{split}
&\overline{\eta}^A(\omega,\zeta) = \frac{2 B_{\text{RB}} N_A \textup{log}_2 e}{(R_U^A)^2-(R_L^A)^2}\int_{R_{\textup{L}}^A}^{R_{\textup{U}}^A}\int_0^{\infty} \frac{1 - \Psi(x,y)}{1+x}y\,\mathrm{d}x\,\mathrm{d}y,
\end{split}
\end{equation}
where
\begin{equation}
   \Psi(x,y) \triangleq e^{-\pi\lambda (R_m^2 - R_{0m}^2) P_A \left[1 - F_{\gamma^A_{u,n}|d_{0,u}}(x|y)\right]}.
\end{equation}

\begin{table}
\renewcommand{\arraystretch}{1.2}
\caption{Network parameters}
\label{t2}
\centering
\scalebox{.9}{
\begin{tabular}{|c|c|}
\hline
\textbf{Parameter} & \textbf{Value}\\
\hline
Cell radius & 500 m\\
Minimum distance between BS and users & 35 m\\
Antenna configuration & SISO\\
Transmit power of the BS & 46 dBm\\
Antenna gain at the BS & 14 dBi\\
Power spectral density of noise & -174 dBm/Hz\\
Receiver noise figure & 7 dB\\
Total bandwidth & 20 MHz\\
Subcarrier spacing & 15 kHz\\
FFT size & 2048\\
Occupied subcarriers (including DC) & 1201\\
Guard subcarriers & 847\\
Number of resource blocks & 100\\
Subcarriers per RB & 12\\
Path loss model & $15.3+37.6\log_{10}(d)$ dB\\
Scheduling policy & proportional fair\\
\hline
\end{tabular}}
\label{tab:parameters}
\end{table}

\section{Numerical results}
\label{sec:Results}
In order to validate the proposed analytical framework while also providing valuable design guidelines, the 19-cell network shown in Fig. \ref{fig:Network_model} is considered, where the cell of interest is surrounded by two rings of interfering BSs. As stated in previous sections, users are distributed over the coverage area using a PPP of normalised intensity $\lambda$. For the sake of presentation clarity, although the analytical framework has been developed using the normalised intensity of the PPP and to ease the visual interpretation, results in this section will be shown as a function of the average number of users per cell $M \triangleq \pi \lambda \left(R_m^2 - R_{0m}^2\right)$. The main system parameters used to generate both the analytical and simulation results have been particularized using numerology drawn from the downlink specification of an LTE/LTE-A network and have been summarized in Table \ref{tab:parameters}. Monte Carlo simulations have been performed using a Matlab\texttrademark-based environment.

\subsection{Background designs}
The proposed R5pD-based design will be compared to optimization designs previously proposed in the literature. In particular, two benchmark designs will be explored:
 \begin{itemize}
\item Fixed spectrum/power factor design (FxD): With this design, the use of a fixed frequency partition is considered when the FFR scheme is employed or a fixed power control when the SFR scheme is used. This design typically leads to high spectral efficiencies at the cost of user fairness \cite{Xu12,Femenias15,garciaperformanceMONET}.
\item Quality-constrained Design (QoScD): Under this design the optimization problem is constrained in order to ensure that the cell-edge throughput is, at least, a fixed fraction of the cell-center throughput. Consequently, a certain fairness degree (i.e., a certain QoS for the cell-edge users) is enforced between cell-center and cell-edge users while maximizing the area spectral efficiency \cite{Jin13,garciaperformanceMONET}.
\end{itemize}

Owing to the fixed spectrum-allocation factor design in FFR, the spectrum allocation factor is selected to be $\rho = \rho_o$ (with, typically, $\rho_o \simeq 1/2$). Consequently, the parameter $\omega$ is the only one that remains to be optimized, thus allowing the optimization problem to be formulated as
\begin{equation}\label{d2}
\omega^* =  \arg\max_{\frac{R_{0m}}{R_m}{\leq\omega\leq}1} \overline{\eta}^C(\omega,\rho_o) + \overline{\eta}^E(\omega,\rho_o).
\end{equation}
When the SFR scheme is used with the fixed power-control factor design, the power factor is fixed to $\beta = \beta_o$ and now, the optimization problem can be formally defined as
\begin{equation}\label{d3}
\omega^* =  \arg\max_{\frac{R_{0m}}{R_m}{\leq\omega\leq}1} \overline{\eta}^C(\omega,\beta_o) + \overline{\eta}^E(\omega,\beta_o).
\end{equation}

In the QoScD approach, a QoS requirement $\varrho$ is specified enforcing that the guaranteed average cell throughput per-edge user is at least a fraction $\varrho$ of the average cell throughput per-center user. Hence, the system parameters are adjusted to trade the data rates required by the cell-center users against those of the cell-edge users. The corresponding optimization problem can be formulated as
\begin{equation}\label{d4}
\begin{split}
(\omega^*,\zeta^*) = & \arg\max_{\substack{\frac{R_{0m}}{R_m}{\leq\omega\leq}1\\\text{ }\text{ }\text{ }\text{ }\text{ }\text{ }0 \leq \zeta \leq 1}} \overline{\eta}^C(\omega,\zeta) + \overline{\eta}^E(\omega,\zeta),\\
                     & \textup{ subject to } \tau_{u}^E(\omega,\zeta) \geq \varrho\, \tau_{u}^C(\omega,\zeta),
\end{split}
\end{equation}
where $\tau_{u}^A(\omega,\zeta) = \frac{\overline{\eta}^A(\omega,\zeta)}{M P_A}$ is the per-user average throughput in cell-region $A$.

\begin{figure*}
        \centering
        \begin{subfigure}[b]{0.31\textwidth}
                \includegraphics[width=\textwidth,height=0.3\textheight]{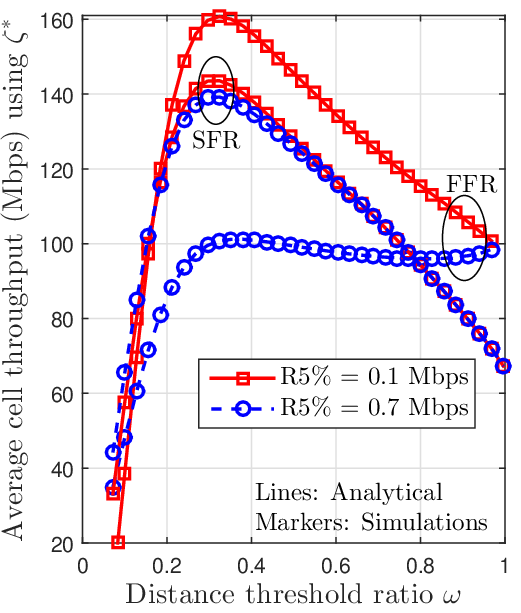}
                \caption{R5pD}
                \label{fig:Thr_vs_w_R5pD}
        \end{subfigure}
        ~~
        \begin{subfigure}[b]{0.31\textwidth}
                \includegraphics[width=\textwidth,height=0.3\textheight]{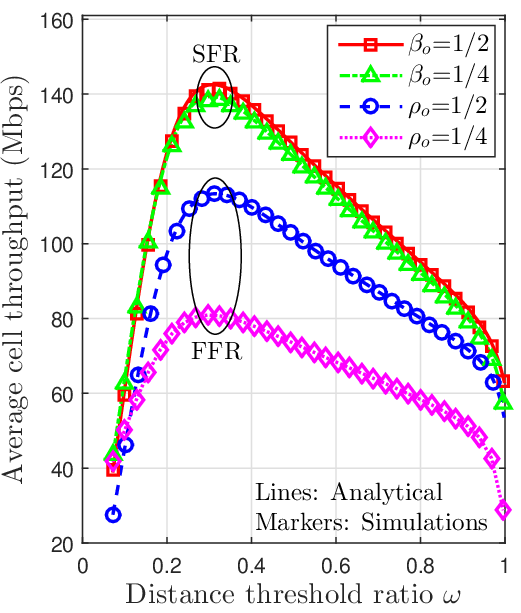}
                \caption{FxD}
                \label{fig:Thr_vs_w_FxD}
        \end{subfigure}
        ~~
        \begin{subfigure}[b]{0.31\textwidth}
                \includegraphics[width=\textwidth,height=0.3\textheight]{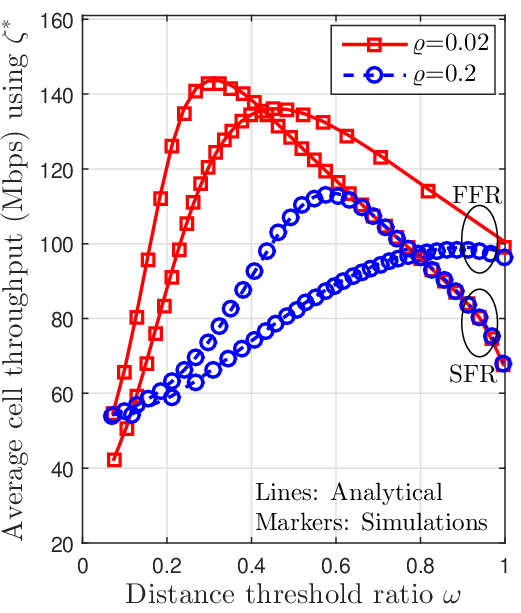}
                \caption{QoScD}
                \label{fig:Thr_vs_w_QoScD}
        \end{subfigure}
        \caption{Average cell throughput versus distance threshold ratio for different optimal designs under both FFR and SFR.}\label{fig:Thr_vs_w}
\end{figure*}

\subsection{Proposed R5pD-based design and Comparison}
Figure~\ref{fig:Thr_vs_w} presents the average cell throughput as a function of the distance threshold ratio $\omega$ under both FFR and SFR reuse schemes.
Illustrating the system behavior for different optimal designs, analytical and simulation results are provided using an average number of users per cell $M=32$. The particular shape of the curves shown in this figure is a direct consequence of the trade-off between the average throughput values attained at both the cell-edge and cell-center regions as a function of the distance threshold ratio $\omega$ and the optimal spectrum or power allocation factor $\zeta^*$. The first point to highlight is the very good match between analytical and simulation results, thus validating the analytical framework that has been developed in Section \ref{sec:Throughput}.

The results, in Fig.~\ref{fig:Thr_vs_w_R5pD}, have been obtained for the R5pD-based design using two different values of the $R5\%$ constraint, corresponding to low and high cell-edge user performance requirements. For each value of $\omega$, the value of $\zeta$ (which is equal to $\rho$ when using FFR and equal to $\beta$ for SFR) maximizing the average cell throughput has been used. The optimal pairs $(\omega^*,\zeta^*)$ in the graphs are indeed the solutions to the problem posed in \eqref{d1}. As expected, increasing the $R5\%$ requirement enforces a higher cell-edge user performance at the cost of decreasing the average cell throughput for both reuse schemes. However, the behavior of both frequency reuse schemes when facing high and low cell-edge user performance requirements is markedly different. First of all, it can be observed that while the average cell throughput of an FFR-based system increases a lot when decreasing the fairness requirement, the average cell throughput of an SFR-based system seems to have very low sensitivity to variations in the 5th-percentile user rate constraint. Furthermore, and related to the previous statement, for a system with a high 5th-percentile user rate constraint (e.g., 0.7 Mbps) the SFR strategy clearly outperforms the FFR scheme, whereas for a system with a low 5th-percentile user rate constraint (e.g., 0.1 Mbps) the SFR scheme is outperformed by the FFR strategy\footnote{Note that, in contrast to what has been considered in our paper, Novlan \emph{et al.} in [9], one of the most cited works in the stochastic geometry-based FFR/SFR analytical frameworks, implicitly assume the use of an over-simplistic round robin (RR) scheduling rule and a uniform distribution of users of sufficient density such that all available sub-bands are utilized, a condition that cannot be guaranteed in realistic deployments serving a reduced number of users and can eventually lead to a non-negligible performance degradation due to the \emph{waste} of resources when allocating sub-bands to unpopulated FFR/SFR-defined cellular regions. However, despite the differences between the assumptions considered in both research works, it is remarkable that the results just shown are consistent with one of the most important results derived in [9]. In particular, as stated in [9, Section V.C], defining $\beta$ as the quotient between the powers allocated to cell-edge and cell-center resources, ``\ldots at low values of $\beta$ [leading to low 5th-percentile user rates], SFR provides lower coverage probability compared to Strict FFR. However, \ldots if $\beta$ is sufficiently large [leading to high 5th-percentile user rates], SFR will surpass strict FFR \ldots''.}. The explanations sustaining these interesting but somewhat counterintuitive results are:
\begin{itemize}
   \item From the results presented in Fig.~\ref{fig:Thr_vs_w_R5pD} it is quite obvious that, for a given average number of users per cell and irrespective of the frequency reuse scheme, the optimal distance threshold ratio $\omega$ is rather insensitive to the variations of the $R5\%$ requirement. Thus, the only parameter playing a central role when optimizing the resource allocation is either $\rho$ in the FFR-based systems or $\beta$ in the SFR-based schemes. Since the network under study is to a large extent an interference-limited cellular system, varying $\beta$ has a very limited effect on the SINR experienced by the served users and, as a consequence, changing the value of $\beta$ has also very limited effects on the average cell throughput. This can be straightforwardly deduced from the mild variations of the optimal average cell throughput the modifications of the $R5\%$ requirement produce. In contrast, adapting the number of subcarriers allocated to both cell-center and cell-edge in accordance to the $R5\%$ constraint can produce large modifications of the average cell throughput. This is because the subcarriers that are not needed to satisfy the throughput requirements of the cell-edge users can be transferred to the cell-center users in order to increase the average cell throughput.

   \item For a high $R5\%$ requirement, the fact that the SFR-based networks reuse the whole set of available subcarriers in each cell of the system, with the consequent increase in potentially achievable spectral efficiency, provides a clear advantage of this frequency reuse scheme in front of FFR. However, for a low $R5\%$ requirement, the FFR-based scheme can take advantage of the flexibility provided by its adaptive frequency allocation capability to increase the average cell throughput while decreasing the throughput experienced by the cell-edge users.
\end{itemize}

For comparative purposes, the average cell throughput obtained when using FxD and QoScD are also presented in Figs.~\ref{fig:Thr_vs_w_FxD} and \ref{fig:Thr_vs_w_QoScD}, respectively. Each value of $\omega^*$ leading to the maximum average cell throughput illustrated in Fig.~\ref{fig:Thr_vs_w_FxD} is indeed the solution to either problem \eqref{d2}, for FFR, or problem \eqref{d3}, for SFR. As the value of $\beta_o$ increases, the optimal average cell throughput of SFR-aided systems improves due to the higher SINRs experienced by the cell-center users, although this comes at the cost of harming the cell-edge throughput due to the increased ICI. In fact, decreasing the power control factor $\beta_o$ causes the cell throughput of SFR to tend to that of FFR for $\rho_o=1/2$. The most important disadvantage of FxD-based schemes in front of strategies based on the proposed R5pD is that the former are unable to adaptively provide QoS guarantees to the users served by the cellular network. Figure~\ref{fig:Thr_vs_w_QoScD} shows the behavior of the QoScD strategy for different QoS constraints. Quality factors of $\varrho = 0.02$ and $\varrho = 0.2$ have been considered, corresponding to low and high fairness requirements between cell-center and cell-edge users, respectively. Irrespective of the applied reuse scheme, increasing the quality factor $\varrho$, and similar to the behavior related to the $R5\%$ constraint, enforces a higher degree of fairness among cell-center and cell-edge users at the cost of decreasing the average cell throughput. Again, for each value of $\omega$, the value of $\zeta$ maximizing the average cell throughput has been used and the pairs $(\omega^*,\zeta^*)$ are now the solutions to problem \eqref{d4}. In this case, QoScD-based schemes can apparently provide similar fairness
related control capabilities as those provided by the R5pD-based strategies. However, note that the QoS constrained design is based on performance metrics that, in contrast to R5pD, are not usually employed in neither the evaluation of current LTE/LTE-A cellular networks nor the specification of envisaged 5G systems.

\begin{figure*}
        \centering
        \begin{subfigure}[b]{0.31\textwidth}
                \includegraphics[width=\textwidth,height=0.3\textheight]{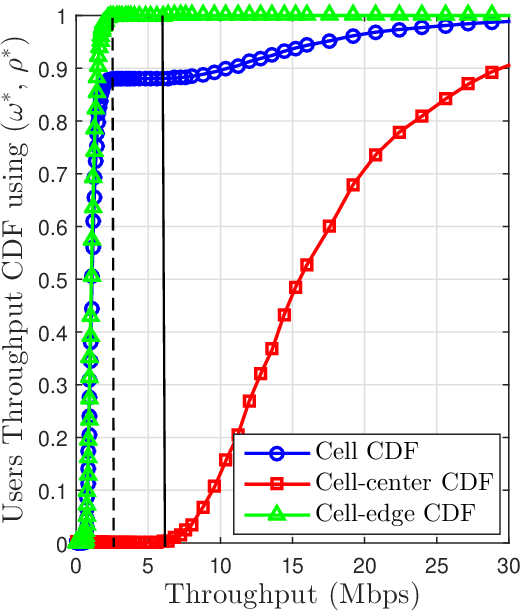}
                \caption{FFR ($R5\%=0.7$ Mbps)}
                \label{fig:CDFr_Regions_R5pD}
        \end{subfigure}
        ~~
        \begin{subfigure}[b]{0.31\textwidth}
                \includegraphics[width=\textwidth,height=0.3\textheight]{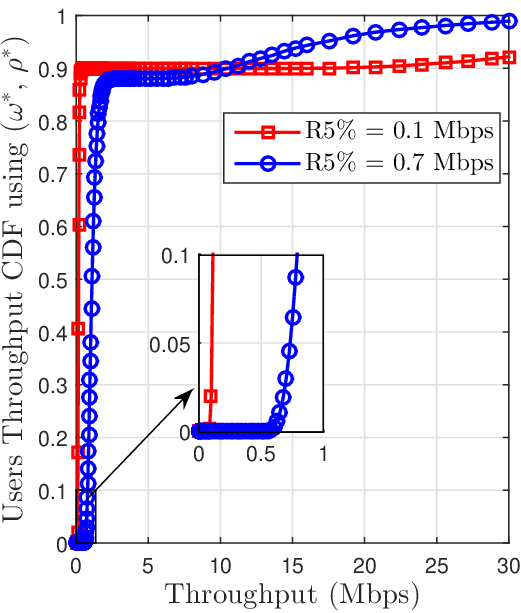}
                \caption{FFR}
                \label{fig:CDFr_R5pD_FFR}
        \end{subfigure}
        ~~
        \begin{subfigure}[b]{0.31\textwidth}
                \includegraphics[width=\textwidth,height=0.3\textheight]{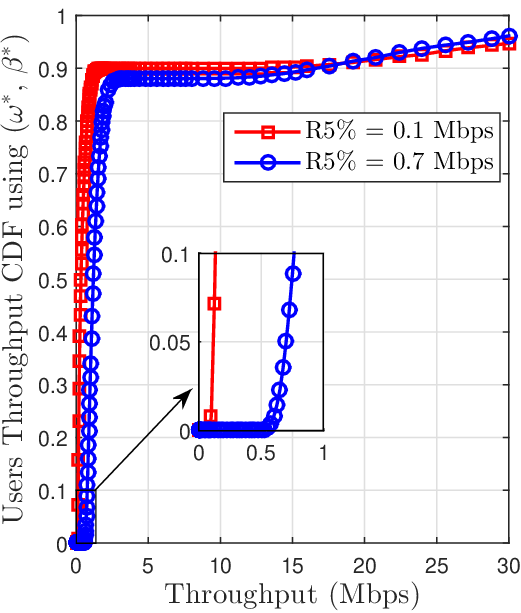}
                \caption{SFR}
                \label{fig:CDFr_R5pD_SFR}
        \end{subfigure}
        \caption{CDF of the per-region user throughput and CDF of the average user throughput for R5pD under both FFR and SFR.}\label{fig:CDF_R5pD}
\end{figure*}

\begin{figure*}
        \centering
        \begin{subfigure}[b]{0.31\textwidth}
                \includegraphics[width=\textwidth,height=0.3\textheight]{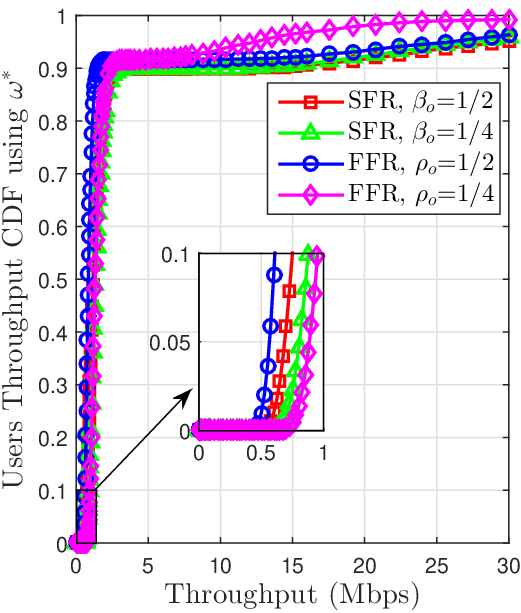}
                \caption{FxD, FFR/SFR}
                \label{fig:CDFr_FxD}
        \end{subfigure}
        ~~
        \begin{subfigure}[b]{0.31\textwidth}
                \includegraphics[width=\textwidth,height=0.3\textheight]{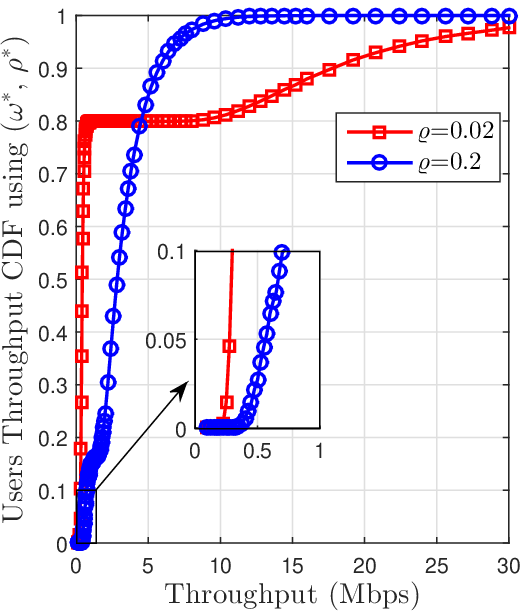}
                \caption{QoScD, FFR}
                \label{fig:CDFr_QoScD_FFR}
        \end{subfigure}
        ~~
        \begin{subfigure}[b]{0.31\textwidth}
                \includegraphics[width=\textwidth,height=0.3\textheight]{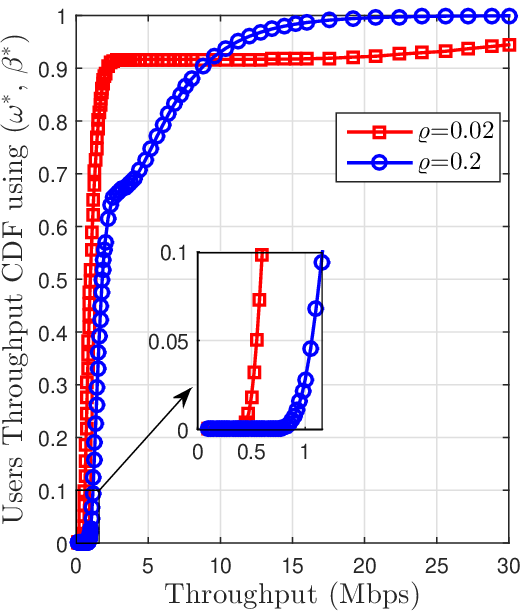}
                \caption{QoScD, SFR}
                \label{fig:CDFr_QoScD_SFR}
        \end{subfigure}
        \caption{CDF of the average user throughput for FxD and QoScD under both FFR and SFR.}\label{fig:CDF_designs}
\end{figure*}

Having established the accuracy of the proposed analytical method and for the sake of clarity, results shown in Fig.~\ref{fig:CDF_R5pD}, depict only the theoretical curves notwithstanding the fact that they have been duly checked via simulation. Figure~\ref{fig:CDFr_Regions_R5pD} shows the CDF of the average user throughput (blue curve with circle markers) for the FFR case and using the constraint $R5\%~=~0.7$~Mbps. Since the pattern observed in this CDF curve will recurrently appear in most graphs shown in this section, it is worth explaining the causes of its shape. To this end, in Fig.~\ref{fig:CDFr_Regions_R5pD} the CDF curves for cell-edge only (green curve with triangle markers) and cell-center only (red curve with square markers) users' throughput are also depicted. The dashed vertical line signals the throughput value for which the cell-edge CDF reaches probability one. That is, there is no cell-edge user experiencing a throughput higher than this value. The solid vertical line pinpoints the throughput value for which the cell-center CDF departs from zero. That is, there is no cell-center user experiencing a throughput lower than this value. Hence, the cell-edge CDF conditions the average user CDF for low throughput values whereas the cell-center CDF conditions the average user CDF for high throughput values. As the throughput value pinpointed by the solid line is higher that the one signaled by the dashed line, the average user throughput CDF shows a plateau whose width is determined by the disparity between the throughput values experienced by cell-edge and cell-center users and its constant value represents the percentage of cell-edge users in the system. Logically, given the uniform user distribution, the percentage of users in each cell region is determined by the optimal distance threshold value (see Fig.~\ref{fig:Thr_vs_w_R5pD}) and thus, the higher the value of $\omega^*$ the lower the level of the plateau.

Figures \ref{fig:CDFr_R5pD_FFR} and \ref{fig:CDFr_R5pD_SFR} represent the CDF of the average user throughput under FFR and SFR, respectively, when considering different values of the $R5\%$ constraint. For the sake of clarity, a zoomed region focusing on the lower values of the curves is also shown where it can easily be checked how the throughput values for which $F_{\overline{\eta}_u}=0.05$ coincide with the corresponding $R5\%$ constraints. The first remarkable point to notice is that as the optimal distance thresholds for SFR are slightly smaller than their FFR counterparts (see Fig.~\ref{fig:Thr_vs_w_R5pD}) and consequently, the plateau values observed for SFR-based networks are somewhat larger than those obtained when using FFR-based frequency reuse schemes. Another interesting fact to highlight is that a lower $R5\%$ requirement leads to a larger disparity between cell-center and cell-edge users' throughput as it can be inferred from the wider plateau width, particularly noticeable for the FFR schemes. Wider plateaus imply larger cell-center throughput in particular and larger average cell throughput in general as it can be corroborated by looking at the results presented in Fig.~\ref{fig:Thr_vs_w_R5pD}. Notice how in SFR-based systems the differences in the plateau widths are much more modest than in FFR-based networks and correspondingly, as it has been mentioned previously, the differences they experience in optimal average cell throughput are barely influenced by the $R5\%$ requirement.

In order to compare R5pD-based results with those obtained under the benchmark designs, Fig.~\ref{fig:CDF_designs} shows the CDF of the average user throughput for FxD and QoScD under both FFR and SFR frequency reuse schemes and assuming an average number of users per cell $M=32$. As shown in Fig.~\ref{fig:CDFr_FxD}, results for FFR/FxD have been obtained using two fixed spectrum allocation factors, namely, $\rho=1/2$ and $\rho=1/4$, and results for SFR/FxD have been obtained using two fixed power allocation factors, namely, $\beta_0=1/2$ and $\beta_0=1/4$. These results reveal, first, that the obtained $R5\%$ values are somewhere in between 0.5 and 1 Mbps, however, these values cannot be controlled by design. Second, in all cases the plateau level is located around 0.9, that is, approximately $90\%$ of the users are located in the cell-edge whose optimal area, as was shown in Fig.~\ref{fig:Thr_vs_w_FxD}, is barely dependent on the frequency reuse strategy (FFR vs SFR) and/or fixed allocation factor ($\rho_0$ or $\beta_0$). Finally, the plateau widths are very sensitive to variations of $\rho_0$ for the FFR-based networks and fairly insensitive to variations of $\beta_0$ for the SFR-based systems, thus corroborating the conclusions extracted from Fig.~\ref{fig:Thr_vs_w_FxD} in which we have stated that the optimal average throughput values in FFR-based networks are very sensitive to changes in $\rho_0$ and they are barely sensitive to changes in $\beta_0$ in SFR-based cellular infrastructures.

Similarly, Figs. \ref{fig:CDFr_QoScD_FFR} and \ref{fig:CDFr_QoScD_SFR} represent the CDF of the average user throughput for FFR/QoScD and SFR/QoScD, respectively, using two quality factors, namely, $\varrho=0.02$ and $\varrho=0.2$. First of all, it can be observed that increasing the quality factor generates an increase in the $R5\%$ value but, again, its particular value cannot be controlled by design. Furthermore, varying the quality factor requirement drastically modifies the shape of the CDF curves, with a large decrease in both the level and the width of the plateau as $\varrho$ increases from 0.02 to 0.2. This serves to corroborate results presented in Fig.~\ref{fig:Thr_vs_w_QoScD}, where it was observed that changing the quality factor produces large variations in both the optimal distance threshold ratio $\omega^*$ and the optimal average cell throughput. Noticeably, the use of high quality factors, which is equivalent to enforcing strong throughput fairness requirements between cell-center and cell-edge users, causes a decrease in the disparity between cell-center and cell-edge users' throughput, which is revealed by the practical disappearance of the plateau in the corresponding CDF curves.

\begin{figure*}
        \centering
        \begin{subfigure}[b]{0.31\textwidth}
                \includegraphics[width=\textwidth,height=0.3\textheight]{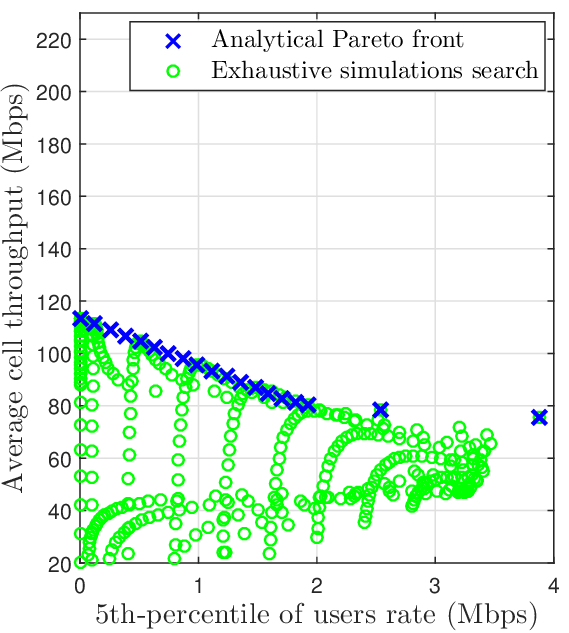}
                \caption{FFR Pareto front, $M = 8$}
                \label{fig:Pareto_FFR_M8}
        \end{subfigure}
        ~~
        \begin{subfigure}[b]{0.31\textwidth}
                \includegraphics[width=\textwidth,height=0.3\textheight]{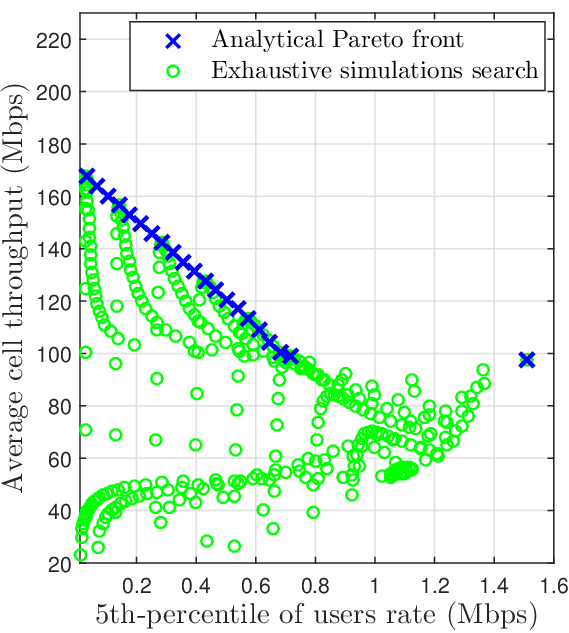}
                \caption{FFR Pareto front, $M = 32$}
                \label{fig:Pareto_FFR_M32}
        \end{subfigure}
        ~~
        \begin{subfigure}[b]{0.31\textwidth}
                \includegraphics[width=\textwidth,height=0.3\textheight]{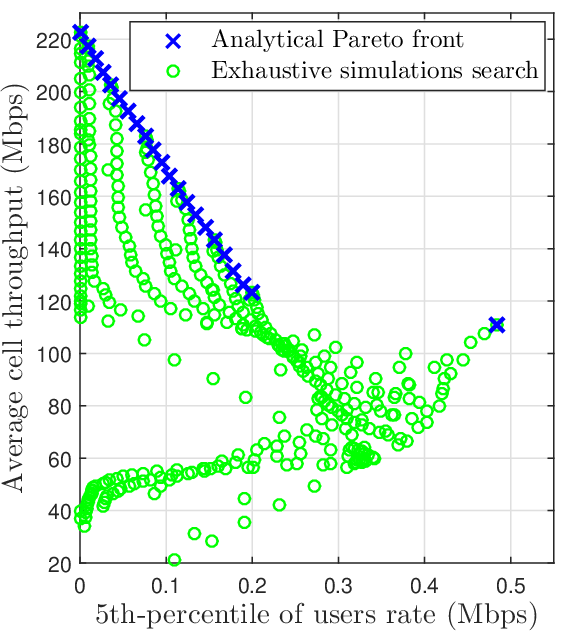}
                \caption{FFR Pareto front, $M = 128$}
                \label{fig:Pareto_FFR_M128}
        \end{subfigure}

        \text{ }

        \begin{subfigure}[b]{0.31\textwidth}
                \includegraphics[width=\textwidth,height=0.3\textheight]{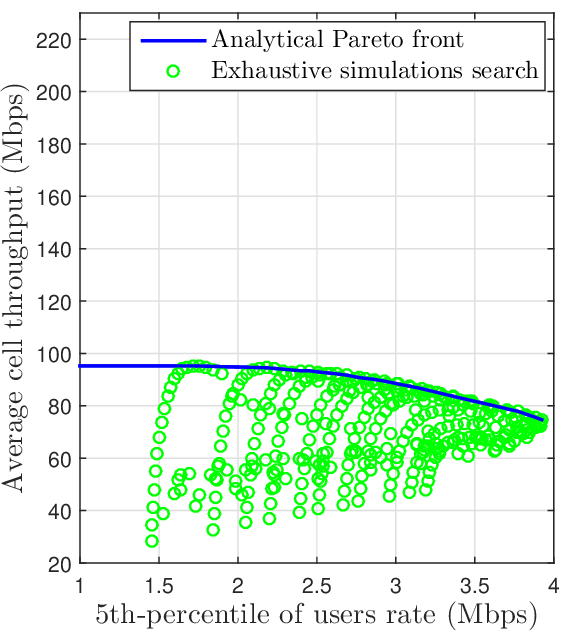}
                \caption{SFR Pareto front, $M = 8$}
                \label{fig:Pareto_SFR_M8}
        \end{subfigure}
        ~~
        \begin{subfigure}[b]{0.31\textwidth}
                \includegraphics[width=\textwidth,height=0.3\textheight]{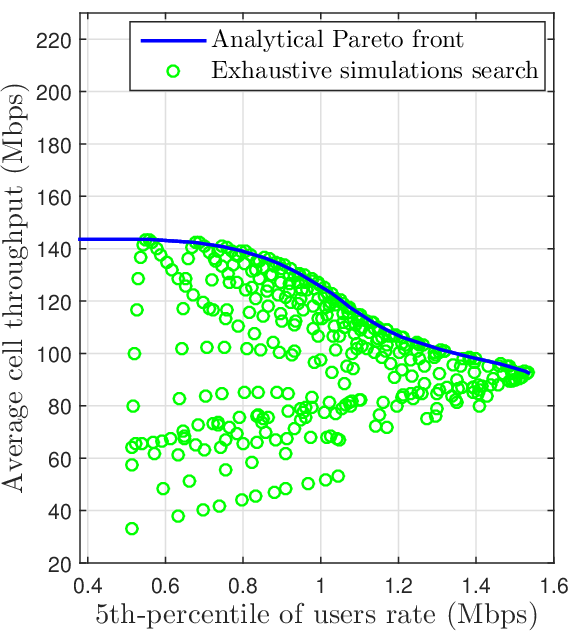}
                \caption{SFR Pareto front, $M = 32$}
                \label{fig:Pareto_SFR_M32}
        \end{subfigure}
        ~~
        \begin{subfigure}[b]{0.31\textwidth}
                \includegraphics[width=\textwidth,height=0.3\textheight]{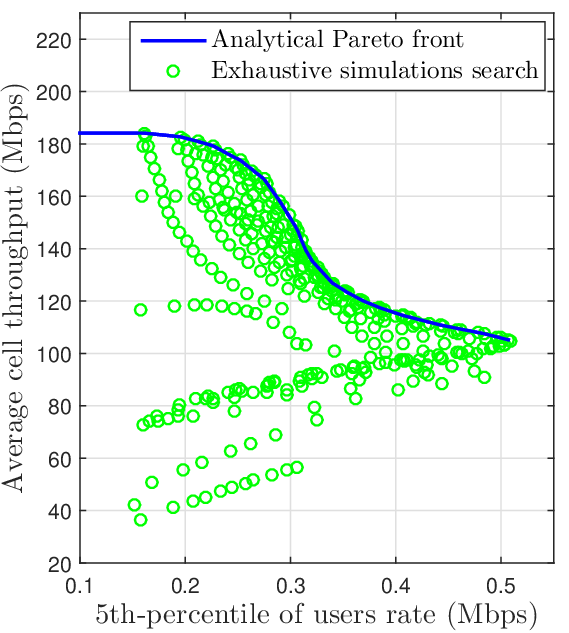}
                \caption{SFR Pareto front, $M = 128$}
                \label{fig:Pareto_SFR_M128}
        \end{subfigure}
        \caption{Average cell throughput versus 5th-percentile user rate for R5pD using different values of $M$ under both FFR and SFR.}\label{fig:Pareto_R5pD}
\end{figure*}

\begin{figure*}
        \centering
        \begin{subfigure}[b]{0.48\textwidth}
                \includegraphics[width=\textwidth]{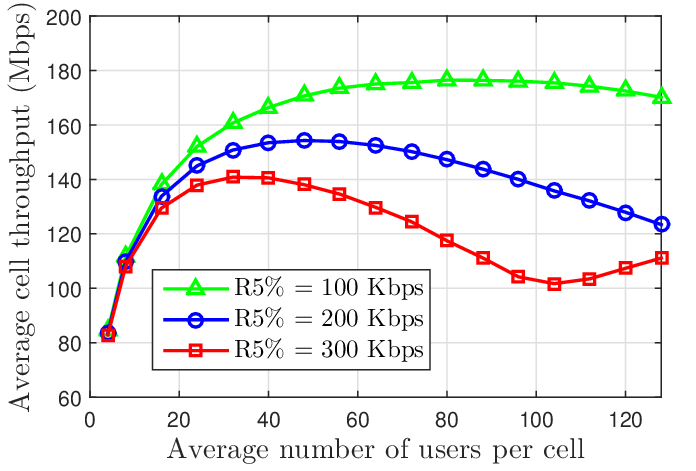}
                \caption{FFR scheme}
                \label{fig:FFR_Thr_vs_M_R5p}
        \end{subfigure}
        ~~
        \begin{subfigure}[b]{0.48\textwidth}
                \includegraphics[width=\textwidth]{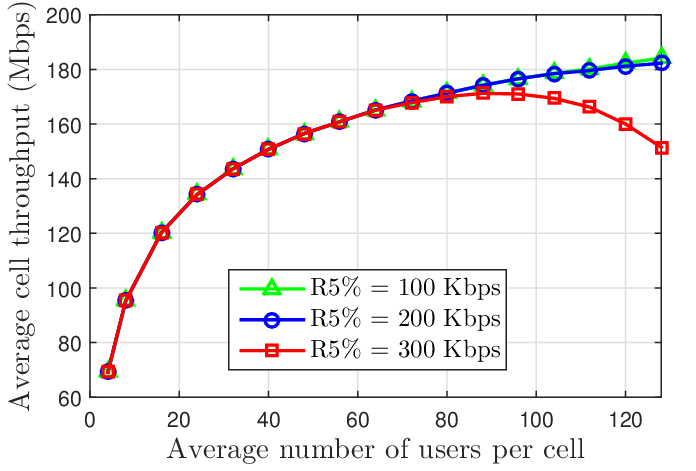}
                \caption{SFR scheme}
                \label{fig:SFR_Thr_vs_M_R5p}
        \end{subfigure}
        \caption{Average cell throughput versus user load using different values of 5th-percentile user rate under both FFR and SFR.}\label{fig:Thr_vs_M_R5p}
\end{figure*}

Figure~\ref{fig:Pareto_R5pD} shows the analytical and simulated average cell throughput as a function of the 5th-percentile user rate constraint considering configurations with different average number of users per cell. The results obtained through simulation correspond to an extensive set of $(\omega,\zeta)$ values, while analytical results are represented only for the optimal pairs $(\omega^*,\zeta^*)$ and assuming, for each one of them, a different $R5\%$ requirement. Using this setup, Figs.~\ref{fig:Pareto_FFR_M8}, \ref{fig:Pareto_FFR_M32} and \ref{fig:Pareto_FFR_M128}, corresponding to FFR, and Figs. \ref{fig:Pareto_SFR_M8}, \ref{fig:Pareto_SFR_M32} and \ref{fig:Pareto_SFR_M128}, obtained for SFR, provide a comprehensive overview of the trade-offs that a network manager can establish among contradictory objective functions such as the average cell throughput and the 5th-percentile user rate. To this end, it is useful to examine these results as solutions to a multiobjective optimization problem (MOP) \cite{Bjornson14} where parameters ($\omega,\zeta$) lead to different performance compromises. A key concept in MOPs is the optimal Pareto front, defined as the set of Pareto optimal solutions (also known as non-dominated, Pareto efficient or non-inferior solutions), defined as those for which none of the objective functions can be improved without degrading some of the other ones. In Fig.~\ref{fig:Pareto_R5pD}, the optimal Pareto front has been obtained using the analytical framework described in previous sections by using, for each particular value of the 5th-percentile user rate, the corresponding optimal pair $(\omega^*,\zeta^*)$. Obviously, analytical solutions representing the Pareto front outperform the rest of solutions obtained by simulation and, remarkably, the optimal simulated values show a very good match with those representing the Pareto front, thus validating once again the analytical framework proposed in this paper. Note that for FFR, a discrete Pareto front has been represented, rather than a continuous one as in the SFR case, because the optimal spectrum allocation factor $\rho^*$ can only take values in the set $\mathcal{S}_\rho$ in \eqref{eq:Srho}. In fact, for the specific system parameters used in the simulation setups only the 5th-percentile of users rate represented by the cross markers are implementable if our aim is to maximize the average cell throughput. For example, for the $M=32$ users case (see Fig. \ref{fig:Pareto_FFR_M32}) setting a target 5th-percentile user rate of, for instance, 1 Mbps provides an average cell throughput of approximately 98 Mbps with a 5th-percentile user rate of roughly 1.54 Mbps, which is obviously higher that the required target value. Note that, due to the discrete set of subcarrier allocations, this is indeed the only point of the Pareto front that actually fulfills the prescribed $R5\%$ target. Furthermore, observe how, in contrast to what happens when using FFR-based frequency reuse schemes, in SFR-based networks there is a wide range of low 5th-percentile user rate values for which the average cell throughput remains virtually constant. That is, decreasing the $R5\%$ requirement in SFR-based networks does not always lead to increasing the global throughput performance of the network and this, as was previously stated, is basically due to that B4G/5G cellular networks are, to a large extent, interference-limited and hence, varying $\beta$ has a very limited effect on the SINR experienced by the served users. This results in the fact, clearly observable when comparing the optimal Pareto fronts depicted in Fig.~\ref{fig:Pareto_R5pD}, that the FFR-based frequency reuse schemes provide a better performance than the SFR-based strategies for low values of the $R5\%$ constraint.

To gain further insight on the interrelation among the average cell throughput, the 5th-percentile user rate and the number of active users in the system, Fig. \ref{fig:Thr_vs_M_R5p} depicts the average cell throughput versus user load under different levels of $R5\%$ for both FFR- and SFR-based networks. Regarding the FFR performance, it can be observed how increasing the $R5\%$ user rate invariably leads to a lower average cell throughput as more resources need to be allocated to the users experiencing poor channel conditions thus decreasing the global throughput performance of the system. Increasing the load, up to a certain number of users, produces an increase in the average cell throughput thanks to the larger degree of multiuser diversity provided by the use of the channel-aware PF scheduler. Beyond this number of users, which decreases when increasing the $R5\%$ user rate, the amount of resources that need to be allocated to the \emph{worst} users in the cell in order for them to achieve the required $R5\%$ clearly masks the benefits produced by the corresponding multiuser diversity and leads to a loss in average cell throughput. Remarkably, as it can be observed in the graph corresponding to $R5\%=300$ Kbps (although not shown in the figure, for lower values of $R5\%$ the same phenomenon would become visible for higher system loads), under very heavy user loads the average cell throughput experiences a non-negligible increase. This can be explained by the fact that the spectrum allocation factor can only take values on the discrete set $\mathcal{S}_\rho$. There is a certain user load for which the $R5\%$ constraint can be fulfilled by using $\rho=(N_{\text{RB}}-3)/N_{\text{RB}}$, that is, there are still 3 RBs left for use in the cell-edge. However, the admission of extra users causes, on one hand, the optimal spectrum allocation factor to become 1, that is, the sudden disappearance of the cell-edge, and on the other hand, the abrupt increase of the optimal 5th-percentile user rate effectively provided to the \emph{worst} users in the system (for instance, we can observe in Fig. \ref{fig:Pareto_FFR_M128} that a system with $M=128$ users with a constraint $R5\%\geq300$ Kbps effectively provides a 5th-percentile user rate of $\sim 490$ Kbps). The elimination of the cell-edge should produce a very large increase in the average system throughput but the increase of $R5\%$ translates to a decrease of this performance metric. The combined effects of these rather radical changes produce the observed moderate increase of the global throughput performance. The effects of user load in SFR-based networks are very similar to those observed in the FFR case except for the fact that, on one hand, due to the interference-limited character of this scheme, the average cell throughput values that can be achieved for light user loads are rather insensitive to the $R5\%$ user rate and, on the other hand, given the continuous nature of the optimization parameters used in the SFR-based networks, heavy user loads do not produce the rather counterintuitive increase of average cell throughput shown by the FFR-based systems. Nonetheless, it can be clearly observed that for a low $R5\%$ requirement and light to moderate user loads, the FFR-based scheme exploits the flexibility provided by its adaptive frequency allocation characteristic to increase the average cell throughput as, in this case, the amount of resources that needs to be allocated to the edge users can be rather limited. For a high $R5\%$ constraint and heavily loaded networks, however, the ability of the SFR scheme to reuse all the subcarriers in each cell provides a clear advantage of this frequency reuse scheme in front of FFR.

\subsection{R5pD-based design guidelines}

\begin{figure}
   \centering
   \includegraphics[width=.96\linewidth]{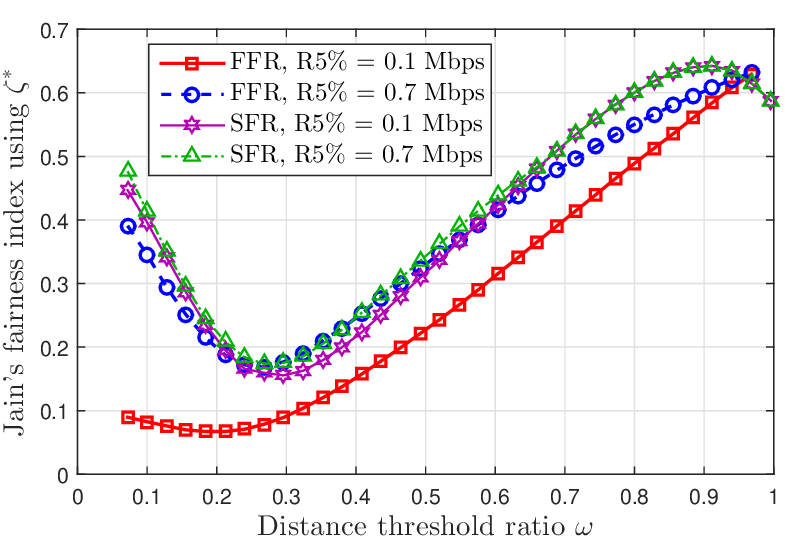}
   \caption{Jain's fairness index of the average user's throughput versus distance threshold ratio for different optimal designs under both FFR and SFR.}\label{fig:Jain}
\end{figure}

Having established the accuracy of the proposed model and in light of the different performance behavior observed for FFR and SFR, it is appropriate to single out some important facts that allow some practical insight to be gained regarding the application of either SFR or FFR:
\begin{itemize}
\item First and foremost, as already mentioned in Section VI.B, under the proposed optimization framework of maximizing overall throughput subject to the fulfilment of a minimum $R5\%$ throughput constraint, the use of FFR is recommended when there are low $R5\%$ requirements, in contrast, SFR is the preferred option for increasing values of the $R5\%$ constraint. Note that this implicitly means that in setups where fairness plays an important role, frequency reuse should take the form of SFR (in alignment with the findings in [9]). In fact, and at a more general level, results shown so far have demonstrated that setting a restriction on the $R5\%$ implicitly serves to balance the cell-center and cell-edge performance.
\item Another important aspect worth taking into account in practical deployments is the performance sensitivity to the different parameters and constraints. In this respect, SFR allows a wider range of $R5\%$ throughput values while barely affecting the overall system performance and, remarkably, the increase of the $R5\%$ constraint induces a graceful degradation of the average cell throughput.
\item Note that $R5\%$ can effectively act as a crank trading off fairness and performance with the former metric being measurable in a variety of manners. Our interest in $R5\%$ is because most wireless communications standards target this metric when setting QoS requirements. Nevertheless, this fact does not preclude the examination of fairness through any other indicator. As an example, Fig. \ref{fig:Jain} shows the Jain's fairness index (JFI) of the throughput among all users in the system with the value of R5\% as parameter. As expected, an increase in the $R5\%$ requirement leads to an increase in the JFI thus confirming an improvement in fairness. In connection with the previous point, SFR is observed to be rather insensitive in terms of JFI whereas FFR is far more influenced by this performance metric.
\item Finally, it is worth mentioning that the discrete nature of the set $\mathcal S_\rho$ in the case of FFR, makes this scheme somewhat more restrictive when it comes to allocating resources to cell-center and cell-edge when compared to the continuous center-edge subcarrier split employed in SFR.
\end{itemize}

\section{Conclusion}
\label{sec:Conclusion}

This paper has presented and validated a novel analytical framework allowing the design of optimal 5th-percentile user rate constrained FFR/SFR-aided OFDMA-based multi-cellular networks using channel-aware PF scheduling. The proposed optimal design, termed R5pD, addresses the maximization of a cell throughput-based utility function subject to a constraint on the 5th-percentile user rate, a typical target requirement in most modern cellular systems. To deal with the R5pD-based optimization problem, an statistical characterization of both per-user and system throughput has been analytically derived and its performance has been thoroughly compared to two previous proposals, namely, the fixed spectrum/power factor design (FxD) and the QoS-constrained design (QoScD).

Regarding the R5pD-based strategy, it has been shown that irrespective of the frequency reuse scheme and in contrast to the benchmark designs, it is able to provide the prescribed 5th-percentile user rate at the cost of sacrificing average cell throughput. Remarkably, it has been observed that the trade-offs at which the FFR- and SFR-based schemes arrive between in terms of system spectral efficiency and user throughput fairness are quite different. For high 5th-percentile user rate constraints, the SFR-based frequency reuse scheme shows a clear advantage when compared to the FFR-based design. This superiority is rooted on the fact that SFR strategies can reuse the whole set of available subcarriers in each cell of the system. When facing low 5th-percentile user rate constraints, however, frequency reuse schemes based on FFR clearly outperform those based on SFR. A rather unexpected behavior arising due to the superior flexibility shown by FFR-schemes in interference-limited cellular networks.

Further work will focus on enhancing the analytical tools developed in this paper to assess the performance of multi-tier cellular networks where a tier of macro-BSs is underlaid by tiers of pico- and/or femto-BSs. Moreover, the framework presented here will be expanded in order to explore new spectrum sharing strategies currently under study in the B4G/5G standardization arena.

\appendix[Statistical distribution of the SINR]
As stated in Section \ref{sec:System_model}, RBs occupy a bandwidth $B_{\text{RB}}$ small enough to assume that all subcarriers in a subband experience frequency flat fading. Moreover,  the proposed scheduling and resource allocation framework relies on capacity formulas optimized on an RB-by-RB basis. Under these assumptions, $h_b=|H_{b,u,n}|^2 $ is subject to an exponential PDF $f_{h_b}(x)= e^{-x} u(x)$ (assuming a Rayleigh fading channel linking the $b$th BS to user $u$ on the $n$th RB), where $u(x)$ denotes the unit step function, and its corresponding CDF expression can be easily obtained as $Pr\{h_b\leq x \}= \left(1-e^{-x}\right) u(x)$. Furthermore, it is not necessary to specify the frequency correlation characteristics of the channel model used in either the analytical framework or the simulation set-up.

The CDF of the instantaneous SINR $\gamma_{u,n}$ conditioned on the set of small-scale fading gains $\boldsymbol{h} \triangleq \{h_b\}_{\forall b \neq 0}$ when the distances $\boldsymbol{d} \triangleq \{d_{b,u}\}_{\forall b}$ from all BSs $b$ are given, can be derived from \eqref{e2} as
\begin{equation}
  \begin{split}
    F_{\gamma_{u,n}|\boldsymbol{d},\boldsymbol{h}}&(x|d,\boldsymbol{h}) \triangleq Pr\{\gamma_{u,n}\leq x | d_{0,u},\boldsymbol{h}\} \\
    &= Pr\left\{h_0\leq \frac{\left(N_0\Delta f + I_{u,n}\right)}{\bar{\gamma}_0} x | d_{0,u},\boldsymbol{h}\right\} \\
    &= 1-e^{-\frac{x\left(N_0\Delta f + I_{u,n}\right)}{\bar{\gamma}_0}}, \qquad x \geq 0,
  \end{split}
\end{equation}
where note that the distances in the set $\boldsymbol{d}$ can be written in terms of the distance $d_{0,u}=d$ from the serving BS to the user $u$ and furthermore, $\bar{\gamma}_0 = P_n L(d)$ represents the average received signal power at a distance $d$ from the tagged BS.

Now, using the above expression and averaging over the PDFs of the i.i.d. random variables $\boldsymbol{h}$, the conditional CDF of the instantaneous SINR $\gamma^A_{u,n}$ experienced by user $u$ located at distance $d_{0,u}=d$ from the serving BS and in the cell-region $A$, for the FFR scheme, can be obtained as
\begin{equation}
  \begin{split}
    &F_{\gamma^A_{u,n}|d_{0,u}}(x|d) \triangleq Pr\{\gamma^A_{u,n}\leq x | d_{0,u}\}\\
    &= \int_0^{\infty}...\int_0^{\infty}\left(1-e^{-\frac{x\left(N_0\Delta f + I_{u,n}\right)}{\bar{\gamma}_0}}\right) \prod_{i\in \Phi_n}f_{h_i}(h_i)\mathrm{d}h_i\\
    &= 1-e^{-\frac{x N_0\Delta f}{\bar{\gamma}_0}}\int_0^{\infty}...\int_0^{\infty}e^{-\frac{x\left(\sum_{i\in \Phi_n} h_i\bar{\gamma}_i\right)}{\bar{\gamma}_0}} \prod_{i\in \Phi_n} e^{-h_i} \mathrm{d}h_i\\
    &= 1-e^{-\frac{x N_0 \Delta f}{\bar{\gamma}_0}}\prod_{i\in \Phi_n} \frac{1}{1 + \frac{x \bar{\gamma}_i}{\bar{\gamma}_0}}\textrm{ }, \qquad x \geq 0,
  \end{split}
\end{equation}
where $f_{h_i}(h_i)$ is the PDF of the variable $h_i=|H_{i,u,n}|^2$, $\overline{\gamma}_0 = P_n L(d)$ represents the average received signal and $\overline{\gamma}_{i} = P_n L\left(d_{i,u}\right)$ is the average interfering signal from each interfering BS.

Under SFR, the conditional CDF of the instantaneous SINR $\gamma^A_{u,n}$ can be expressed as
\begin{equation}
\begin{split}
F_{\gamma^A_{u,n}|d_{0,u}}(x|d)= &1-e^{-\frac{x N_0 \Delta f}{\bar{\gamma}_0}}\prod_{i\in \Phi_n^C} \frac{1}{1 + \frac{x \bar{\gamma}_i}{\bar{\gamma}_0}} \\
                                 &\times\prod_{j\in \Phi_n^E} \frac{1}{1 + \frac{x \bar{\gamma}_j}{\bar{\gamma}_0}}, \qquad x \geq 0,
\end{split}
\end{equation}
where, now, the average received signal is $\overline{\gamma}_0 = P_n^A L(d)$, and the average interfering signals are $\overline{\gamma}_{i} = P_n^C L\left(d_{i,u}\right)$ and $\overline{\gamma}_{j} = P_n^E L\left(d_{j,u}\right)$. Again, $d_{i,u}$ and $d_{j,u}$ can be written in terms of $d$.

\section*{Acknowledgements}

This work has been supported in part by the Agencia Estatal de Investigaci\'on and Fondo Europeo de Desarrollo Regional (AEI/FEDER, UE) under project ELISA (subproject TEC2014-59255-C3-2-R), Ministerio de Economia y Competitividad (MINECO), Spain, and the Conselleria d'Educaci\'o, Cultura i Universitats (Govern de les Illes Balears) under grant FPI/1538/2013 (co-financed by the European Social Fund).

\bibliographystyle{IEEEtran}
\bibliography{biblio}

\end{document}